# Strain-engineering of the charge and spin-orbital interactions in Sr$_2$IrO$_4$

Eugenio Paris[*1], Yi Tseng[1], Ekaterina M. Pärschke[2,3], Wenliang Zhang[1], Mary H. Upton[4], Anna Efimenko[5], Katharina Rolfs[6,7], Daniel E. McNally[1], Laura Maurel[8,9], Muntaser Naamneh[1], Marco Caputo[1], Vladimir N. Strocov[1], Zhiming Wang[10,11], Diego Casa[4], Christof W. Schneider[9], Ekaterina Pomjakushina[6], Krzysztof Wohlfeld[12], Milan Radovic[1] & Thorsten Schmitt[*1]

[1]Photon Science Division, Paul Scherrer Institut, 5232 Villigen PSI, Switzerland
[2]Department of Physics, University of Alabama at Birmingham, Birmingham, Alabama 35294, USA
[3]Institute of Science and Technology Austria, Am Campus 1, 3400 Klosterneuburg, Austria
[4]Advanced Photon Source, Argonne National Laboratory, Argonne, Illinois 60439, USA
[5]ESRF – The European Synchrotron, 71, avenue des Martyrs, CS 40220, 38043 Grenoble, France
[6]Neutrons and Muons Research Division, Paul Scherrer Institut, 5232 Villigen PSI, Switzerland
[7]Physikalisch-Technische Bundesanstalt (PTB) Berlin, Department 8.2 Biosignals, 10587 Berlin, Germany
[8]Laboratory for Mesoscopic Systems, Department of Materials, ETH Zurich, 8093 Zurich, Switzerland
[9]Laboratory for Multiscale Materials Experiments, Paul Scherrer Institute, 5232 Villigen PSI, Switzerland
[10]Key Laboratory of Magnetic Materials and Devices, Ningbo Institute of Materials Technology and Engineering, Chinese Academy of Sciences, Ningbo 315201, Peoples Republic of China
[11]Zhejiang Province Key Laboratory of Magnetic Materials and Application Technology, Ningbo Institute of Materials Technology and Engineering, Chinese Academy of Sciences, Ningbo 315201, China
[12]Institute of Theoretical Physics, Faculty of Physics, University of Warsaw, Pasteura 5, PL-02093 Warsaw, Poland

* Corresponding authors: Eugenio Paris & Thorsten Schmitt

**E-mail**: eugenio.paris@psi.ch; thorsten.schmitt@psi.ch

**Author Contributions**

M. R. and T. S. designed research. E. Pa., Y. T., E. M. P., W. Z., M. H. U., A. E., K. R., D. E. M., L. M., M. N., M. C., V. N. S., Z. W., D. C., C. W. S., E. Po., K. W., M. R., T. S. performed research. E. Pa. and T. S. analyzed the data. E. Pa., E. M. P., K. W., T. S. wrote the paper with input from all authors. Y.T and E. M. P. contributed equally to this work.

**Competing interest statement**

The authors declare no competing interest.

**Abstract**

In the high spin-orbit coupled $Sr_2IrO_4$, the high sensitivity of the ground state to the details of the local lattice structure shows a large potential for the manipulation of the functional properties by inducing local lattice distortions. We use epitaxial strain to modify the Ir-O bond geometry in $Sr_2IrO_4$ and perform momentum-dependent Resonant Inelastic X-ray Scattering (RIXS) at the metal and at the ligand sites to unveil the response of the low energy elementary excitations. We observe that the pseudospin-wave dispersion for tensile-strained $Sr_2IrO_4$ films displays large softening along the [h,0] direction, while along the [h,h] direction it shows hardening. This evolution reveals a renormalization of the magnetic interactions caused by a strain-driven crossover from anisotropic to isotropic interactions between the magnetic moments. Moreover, we detect dispersive electron-hole pair excitations which shift to lower (higher) energies upon compressive (tensile) strain, manifesting a reduction (increase) in the size of the charge gap. This behavior shows an intimate coupling between charge excitations and lattice distortions in $Sr_2IrO_4$, originating from the modified hopping elements between the $t_{2g}$ orbitals. Our work highlights the central role played by the lattice degrees of freedom in determining both the pseudospin and charge excitations of $Sr_2IrO_4$ and provides valuable information towards the control of the ground state of complex oxides in the presence of high spin-orbit coupling.

**Main Text**

**Introduction**

In strongly correlated electron materials, the intricate entanglement of different degrees of freedom gives rise to many interesting phases and emerging phenomena such as metal-insulator transitions, colossal magnetoresistance, and superconductivity to mention a few (1–3). The intertwined degrees of freedom, which often compete, in certain cases cooperate to establish the ground state. A prominent example for this is $Sr_2IrO_4$, where the large spin-orbit coupling combined with the moderate Coulomb repulsion opens a Mott-like gap in the nominally half-filled Ir *5d* $t_{2g}$ orbitals, leading to an unexpected insulating behavior (4, 5).

Such a ground state, consisting of a single hole effectively carrying a spin-orbital total angular momentum (pseudospin) j = ½ in a square lattice, is reminiscent of the high-$T_c$ Cu-oxides. The related prediction of superconductivity (6) has fueled numerous studies in the recent years, which have found impressive similarities between the two families of materials such as the presence of pseudogap (7–10), d-wave symmetry (11, 12), electron-boson coupling (13), and analogies in the magnetic excitations (14, 15). Nonetheless, superconductivity is still elusive in $Sr_2IrO_4$.

Since the ground state properties are a direct consequence of the delicate balance between interwoven degrees of freedom, understanding this intricacy and finding appropriate tuning knobs promises novel functionalities. In this regard, a peculiar characteristic of $Sr_2IrO_4$ is the unprecedented dependence of the physical properties on the local lattice structure in the ground state (16–18). Indeed, due to the entanglement of spin and orbital degrees of freedom, the local structural details play a central role in stabilizing the canted antiferromagnetic (AF) ground state (19, 20).

$Sr_2IrO_4$ adopts the perovskite structure shown in Fig. 1(a). Upon doping or temperature, the local Ir-O bond lengths change only marginally (21), while the Ir-O-Ir bond angle responds clearly (17, 22). In addition, large changes in the bond angle have been observed upon applying electrical currents (23). In all these cases, a dramatic evolution of the ground state properties accompanies the change in the lattice structure. Such findings leave open the question on the role of the lattice, separated from other contributions such as band filling or disorder, urging for an approach involving direct manipulation of the bond geometry. Efforts in this direction provided promising results in view of controlling the ground state properties. For instance, the application of hydrostatic pressure quenches the local magnetic moment of $Sr_2IrO_4$ (24) and induces a magnetic crossover, characterized by the renormalization of the exchange couplings (25). Besides external pressure, epitaxial strain - achieved by growing thin films of $Sr_2IrO_4$ on substrates with mismatching in-plane lattice parameter - allows exerting very high biaxial pressure along the Ir-O bond network in the basal plane (26, 27), as shown schematically in Fig. 1(c). Epitaxial strain can also affect the Ir-O-Ir bond angle, on which the hierarchy between isotropic and anisotropic pseudospin interactions is expected to depend (16). In previous reports, a decrease in the zone-boundary magnon(28) and bimagnon (29) energy was found upon tensile strain, accompanied by an increased Neel temperature (28). Moreover, theoretical proposals suggest the Mott gap to shrink upon compressive strain, driving the system towards an insulator-to-metal transition (30).

Due to its giant anisotropic magnetoresistance, $Sr_2IrO_4$ is a promising material in the novel field of AF spintronics (31–33). For such applications, it is pivotal to be able to control the magnetic interactions via strain in few-nm thin films in the prospect of integrating such material layers into actual devices. In this regard, it is important to obtain information on the evolution of microscopic electronic quantities and the related low-energy elementary excitations in response to the lattice tuning. Such information is also highly valuable towards a better understanding of the magneto-elastic coupling (16, 18, 23, 34, 35) and electron-phonon coupling(13, 36, 37) in $Sr_2IrO_4$.

In this work, we investigate the strain effects on the spin-orbital and charge excitations of $Sr_2IrO_4$ using a combination of O K-edge and Ir $L_3$-edge Resonant Inelastic X-ray Scattering (RIXS). We find the local bond geometry to control the nature of the pseudospin exchange, effectively modulating the degree of anisotropy of the magnetic interactions. The modification of the magnetic dispersion is accompanied by a continuous evolution of the low-energy electronic structure, revealed by a large change in the electron-hole pair excitations. In particular, the strain-induced enhancement of the electron hopping strength causes a renormalization in the size of the Mott gap.

**Results and discussion**

In recent years, intense efforts have been made to exploit the electron-lattice coupling to induce new functionality in $Sr_2IrO_4$. An important missing piece of knowledge in this regard is the study of the collective elementary excitations upon tuning the lattice parameters, which can only be probed using advanced spectroscopies such as RIXS. So far, Ir $L_3$-edge RIXS has been extensively employed to study iridium oxide materials in the bulk form (14, 15, 38–47). It was recently reported that, due to the large metal-ligand hybridization combined with high spin-orbit coupling at the metal ion, information on the elementary excitations can be extracted using an indirect RIXS process at the O K-edge (48). One of the advantages of this approach is the lower penetration depth of soft x-rays, providing high sensitivity to ultrathin samples such as films.

Figures 2 (c) and (d) display the Ir $L_3$-edge and O K-edge RIXS spectra, respectively, both measured at the in-plane momentum transfer Q = (1/4,0) r.l.u. on a $Sr_2IrO_4$ thin film with ε = -0.5% compressive strain. In the O K-edge RIXS experiment, we tune the incoming photon energy to a particular absorption peak, allowing to isolate the contribution to the RIXS signal from the planar oxygen $2p_x/2p_y$ orbitals, hybridized with the Ir $5d$ orbitals (SI Appendix, Fig. S3(b)). In the Ir $L_3$-edge RIXS experiment, instead, the incoming photon energy is tuned to the $L_3$ x-ray absorption resonance, corresponding to the Ir $2p_{3/2}$-$5d$ transition and being directly sensitive to the elementary excitations involving the Ir $d$-shell. Both the Ir $L_3$ and O K-edge RIXS spectra show single j = 1/2 pseudospin-flip excitations centered around 150 meV energy loss [labeled "M" in Fig. 2]. The high sensitivity to spin excitations is due to the presence of strong spin-orbit coupling in the intermediate state in the former case (49) and sizable spin-orbit coupling in the valence band in the latter (48). The magnetic excitations are more intense in the $L_3$-edge spectrum while broader and less defined in the O K-edge spectrum due to the presence of a large bi-(multi)-magnon tail. Both spectra show an intense structure from 0.5 to 1 eV energy loss due to orbital excitations between j = 3/2 and j = 1/2 manifolds (14, 42). On the other hand, the main qualitative difference between the Ir $L_3$-edge and O K-edge RIXS spectrum is the higher sensitivity of the latter to an excitation located at ≈ 400 meV energy loss [labeled "A" in Fig. 2]. As we will discuss later, this excitation is sensitive to the carrier dynamics through electron-hole pair excitations.

Before discussing the evolution of the elementary excitations probed by RIXS, we briefly discuss the possible effects of epitaxial strain on the local structure of $Sr_2IrO_4$. In perovskite oxide materials, the epitaxial strain is expected to induce rotations of the metal-oxide octahedra [see Fig.1 (c)]. Namely, tensile (compressive) strain tends to increase (decrease) the Ir-O-Ir bond angle (26, 27). However, to what extent this rotation would be "rigid" in $Sr_2IrO_4$ is not clear, as a concomitant modulation of the Ir-O bond length is likely at play (28, 30, 50).
In the following, we will first address the effect of strain on the isospin excitations and then discuss the evolution of the low-energy band structure by analyzing the electron-hole pair excitations. The orbital excitations are also modified by strain, a brief discussion on this aspect can be found in the SI Appendix.

**Tuning the magnetic interactions.** The momentum dependence of the Ir $L_3$-edge RIXS spectrum for a $Sr_2IrO_4$ thin film with ε = -0.5% compressive strain is presented in Fig. 3(a). In this sample, as well as in all others investigated in this study, the collective magnetic excitation disperses away from the AF zone center and reaches the band top at the AF zone boundaries, in good agreement with reports on single crystals of $Sr_2IrO_4$(14, 38, 40, 46). In Fig. 3 (b) and (c), we present the evolution of the magnetic excitation as a function of strain at the zone boundaries (1/2,0) and (1/4,1/4), respectively. The magnetic excitation shifts to lower energies upon tensile strain at **Q** = (1/2,0), while an opposite behavior is observed at **Q** = (1/4,1/4), where the magnetic mode hardens. An anisotropic response of the magnetic excitations is reminiscent of electron doping, affecting mainly the [h,h] direction in La-doped $Sr_2IrO_4$ (38, 40, 46) and $Sr_3Ir_2O_7$ (45, 47). It is worth noting that the present anisotropic evolution is dissimilar to the case of a Mott insulator such as $La_2CuO_4$, in which epitaxial strain affects the magnetic dispersion in a similar way along the nodal and antinodal directions (51). As summarized in Fig. 3(d), upon expanding the lattice structure to ε = 1.5%, the magnetic modes along the [h,0] and [h,h] directions of the AF-zone tend to reach a similar energy scale. In order to fully probe the reciprocal lattice, we combine RIXS measurements at the O K and Ir $L_3$-edge to uncover the full dispersion of the magnetic excitations. For each sample, we performed at least one measurement at both O K and Ir $L_3$-edge at a given Q-vector, to ensure the compatibility between the dispersion obtained from the two datasets. The different components in the RIXS spectra are fitted using multiple peak functions. A typical fit obtained for O K-edge and Ir $L_3$ edge at the same exchanged wave vector on a $Sr_2IrO_4$/LSAT film (*ε* = -0.5 %) is presented in Fig. 4(e). The magnetic dispersion, obtained for different strain levels, is presented in Fig. 4 (a)-(d). When the strain level is small (*ε* = ± 0.5 %), the spin-wave dispersion resembles the one observed in the single crystal (14, 40), reaching a band top of ~ 200 meV at Q = (1/2,0) and ~ 100 meV at Q = (1/4,1/4). For higher compressive strain (*ε* = - 0.7 %) the dispersion shows a reduced curvature along the [h,0] direction and tends to reach higher energy at the zone boundary. Upon large tensile strain (*ε* = + 1.5 %) the dispersion presents an increased curvature and a large softening at Q = (1/2,0), with a band top of ~ 160 meV. The softening is accompanied by a hardening at Q = (1/4,1/4), where the magnetic mode reaches ~ 130 meV.

In order to capture the evolution of the exchange couplings upon strain, we fit the spin-wave dispersion using the anisotropic Heisenberg model with longer-range spin interactions, which was recently used to model the magnetic excitations in a bulk $Sr_2IrO_4$ crystal (38, 52) (SI Appendix, eq. S3). The best fit, obtained for different strain levels, is shown as a solid black line in Fig. 4 (a)-(d) and the obtained exchange couplings are reported in Table 1. By increasing the tensile strain, the first-($J_1$), second- ($J_2$), and third nearest neighbor ($J_3$) exchange coupling parameters clearly evolve. The relatively large decrease of the longer-range spin exchanges ($J_2$ and $J_3$) upon increase of the tensile strain is consistent with an associated decrease of the hopping parameters, as pointed out in recent Raman investigations(29). On the other hand, the nearest-neighbor spin exchange $J_1$ experiences relatively small but non-monotonic changes with the increase of the tensile strain.

Figure 4(f) shows the dependence on strain of the ratio of the second and third neighbor exchange interactions with respect to the first (nearest) neighbor exchange ($|J_2|+J_3$)/$J_1$. When $\varepsilon$ = -0.7%, the longer-range spin interactions reach up to 50% of the nearest neighbor exchange. By applying tensile strain, this ratio decreases linearly and, when $\varepsilon$ = +1.5 %, the spin interactions evolve towards a nearest-neighbor Heisenberg-like behavior, with longer-range correlations accounting only for ≈ 15% of $J_1$. In this regard, tensile strain makes the isospin-wave dispersion similar to the spin-wave dispersion observed in cuprates. For comparison, in correlated cuprate insulators such as $La_2CuO_4$, the importance of longer-range terms is ≈ 10% (53).

It has been proposed that a Mott insulator, with strong on-site spin-orbit coupling and metal-oxygen-metal bond close to 180°, can be described with a magnetic Hamiltonian composed by Heisenberg-like interactions with a dominant nearest-neighbor exchange ($J_1$). Such a model contains only small anisotropic corrections, and the magnetic dispersion is expected to be similar along the [h,0] and [h,h] directions of the reciprocal lattice. In $Sr_2IrO_4$ however, the presence of octahedral rotations enhances the importance of Dzyaloshinskii–Moriya and other anisotropic interactions (16). The resulting magnetic dispersion of $Sr_2IrO_4$ shows a different behavior along the [h,0] and [h,h] directions, requiring to include sizable second and third nearest-neighbor exchange interactions ($J_2$ and $J_3$) in the model calculations. Moreover, in most 3$d$ Mott insulators, the AF order arises among localized S=1/2 spins, naturally allowing to map the spin interactions into a Heisenberg model with local interactions. In $Sr_2IrO_4$ on the other hand, the delocalized nature of the 5$d$ orbitals suggests describing the ground state in an intermediate Mott-Slater regime, i.e. to assume a moderate value of the Coulomb repulsion (as compared to the kinetic energy of the electrons) (54). Importantly, the values of $J_2$ and $J_3$ obtained experimentally for bulk single crystals of $Sr_2IrO_4$ are as large as 33% and 25% of $J_1$ in magnitude, respectively. The non-local interactions are thus implicitly included in the phenomenological model through the large higher-order exchange terms. Crucially, the $J_1$-$J_2$-$J_3$ Heisenberg model with longer-range spin interactions employed here describes remarkably well the observed experimental pseudospin-wave dispersion (14, 38, 40) which justifies its choice in this work.

**Tuning the low-energy electronic structure.** In the following, we address the nature of peak A in the O K-edge RIXS spectrum. Through this mode, we can track the evolution of the low-energy electronic structure of $Sr_2IrO_4$ upon misfit strain. Figure 5(a) shows the O K-edge RIXS spectra of a thin film of $Sr_2IrO_4$ with $\varepsilon$ = -0.5%, measured by varying the exchanged momentum in the reciprocal space. Peak A appears as a dispersive mode with a minimum located at the zone center. This excitation has been observed in various iridate compounds (40, 42, 55) but its nature is not well understood. The energy scale (300-400 meV) is of the order of magnitude of the on-site Coulomb interaction and similar to the optical gap energy (26, 36, 56), suggesting that this excitation is related to the formation of electron-hole pairs. Indeed, this excitation is reported to be in the same energy range also in $Na_2IrO_3$ (55), having a band gap similar to $Sr_2IrO_4$ and absent (in this energy range) in $Sr_3Ir_2O_7$ having a smaller charge gap (48). Theoretical work suggested it results from the interaction between optical electron-hole excitations and the spin-orbit exciton modes (57). Moreover, previous RIXS reports ascribed it to an additional branch of the spin-orbit exciton (40), possibly involving a Jahn-Teller like distortion in the $t_{2g}$ orbitals (58).

To clarify the origin of this excitation, we compute the RIXS intensity for an electron-hole inter-band transition by evaluating the momentum resolved joint density of states N($\omega$,**Q**), using the following formula (59):

$$N(\omega, \boldsymbol{Q}) = \sum_v \int_V d\boldsymbol{k} \int_{V'} d\boldsymbol{k}' \delta\big(\hbar\omega - \epsilon_v(\boldsymbol{k}) - \epsilon_c(\boldsymbol{k}')\big) \delta(\boldsymbol{k} - \boldsymbol{k}' - \boldsymbol{Q}) \qquad (1)$$

where $\epsilon_v$ and $\epsilon_C$ represent the valence (occupied) and conduction (unoccupied) bands involved in the electron-hole pair formation. In our notation, $\omega$ is the energy lost by the photon in the RIXS process and **Q** is the exchanged momentum. Since this calculation requires the knowledge of $\epsilon_v(\boldsymbol{k})$ and $\epsilon_C(\boldsymbol{k})$ across the whole reciprocal space, we calculate the dispersions of the electron removal and addition states using single-particle Green functions, accounting for electron-electron correlations in a non-perturbative way, as was recently used to describe photoemission and inverse photoemission spectral functions, respectively (60) (see SI Appendix for details). Finally, we account for the experimental resolution adding a finite Gaussian broadening to the obtained spectrum. The resulting momentum and energy-dependent spectral function for the low-energy electron-hole pair excitations (EHP) in $Sr_2IrO_4$ is presented in Fig. 5(b). In our calculation, when the energy loss equals to the indirect charge gap at ~ 250 meV (7), we observe a continuum in the $N(\omega,\mathbf{Q})$ due to the onset of electron-hole transitions. On top of such a broad structure, we identify a dispersive mode featuring a U-shaped dispersion, with a bottom located at $\mathbf{Q}= (0,0)$. Our calculation captures well the overall behavior of peak A, allowing us to assign it to an interband EHP. The value of the energy minimum at the zone center provides information on the optical gap, for which experimental reports range from ~ 300 meV (26, 36) to 550 meV (56). It is worth noting that the onset of the EHP overlaps with the many-body magnetic excitations, located between 200-300 meV, indicating a possible interaction between charge and isospin degrees of freedom.

To address the effect of strain on the electronic structure of $Sr_2IrO_4$, we collect momentum-dependent O K-edge RIXS on our series of strained thin films. Figure 5(c) shows the strain-dependence of peak A, measured near $\mathbf{Q} = (0,0)$. As one can see, the strain directly affects the energy of this mode, suggesting the underlying modification of the low-energy electronic structure. In particular, peak A hardens in energy upon tensile strain. We analyzed the dependence of the excitation energy on strain and momentum by a detailed peak fitting [see, for example, Fig. 4(e)]. The momentum dispersion of peak A upon various strain values is shown in Fig. 5(e). Regardless of the strain level, this excitation shows a gap at the zone center and disperses, rather symmetrically, along the [h,0] and [h,h] directions. The strain appears to modulate the electron-hole bandwidth, and a large increase in the size of the gap at $\mathbf{Q} = (0,0)$ is observed upon tensile strain. We note that the hardening of peak A resembles the evolution of the "α-peak" observed in optical spectroscopy that was associated with a reduction of the U/t ratio (26, 61). In particular, Nichols *et al.* found the optical gap to increase (decrease) upon tensile (compressive) strain (26). The present findings are in agreement with this scenario and suggest that the material is evolving towards a metallic state upon compressive strain. We observe that going from a compressive strain of -0.7% to a tensile strain of +1.5% we detect a change in the direct charge gap of about ~60 meV. For comparison, upon hydrostatic pressure as large as 30 GPa, the change in the insulating gap inferred from transport measurements was ~30 meV (24). This finding shows the effectiveness of epitaxial strain in tuning the electronic structure by distorting the local Ir-O environment. Nonetheless, the gap remains sizable as a function of strain, allowing us to employ the same approach, based on the localized limit, to describe the charge and spin excitations.

In order to understand the evolution of the EHP excitations microscopically, we compute the joint density of states $N(\omega,\mathbf{Q})$ for different strain levels using Eq. (1). To account for the strain effects, we *firstly* assume an exponential scaling of the orbital-dependent hopping parameters as a function of the in-plane lattice constant, as in (62). *Secondly*, we adapt the exchange coupling parameters to those obtained from fitting the experimental magnetic dispersion for various strain levels, as shown in Fig. 4 (a-d) (details in the SI Appendix). The result of this calculation is shown in Fig. 5(d) at the same momentum transfer vector $\mathbf{Q}$ as the data in Fig. 5(c). The model captures well the evolution of the excitation gap upon strain, i.e. the charge gap increases with increasing ε. In this material, the size of the Mott gap depends not only on the on-site Coulomb repulsion U, but also on the binding energies of the so-called magnetic polarons, which are composed by charge carriers coupled to the magnetic J=1/2 excitations (see SI Appendix). Since U is a local interaction, we assume that this would change only negligibly in the considered range of strain. Therefore, we ascribe the observed increase of the charge gap upon tensile strain to a decrease in the binding energy of magnetic polarons in both valence and conduction bands. Such binding energy depends on the ratios of the spin-exchange parameters J with respect to the hopping parameters t (63). In particular, the exchange parameters $J_2$ and $J_3$ are responsible for the free hopping of the magnetic polaron and contribute mostly to its bandwidth. Upon tensile strain, $J_2$ and $J_3$ decrease quicker than the coupling of the holes and electrons to the magnetic background (~ t) (63), causing the decrease of polaron binding energies resulting in the effective tuning of the charge gap.

As discussed earlier, a reduction of the lattice parameter corresponds locally to a reduction of the Ir-O bond length and/or an enhanced rotation of the $IrO_6$ octahedra. While a rise in the hopping parameters follows naturally a decrease in the Ir-O bond length, such a development is not obvious for an increase in the octahedral rotation. Indeed, it apparently contradicts the existing paradigm in transition metal oxides, that the closer the metal-oxygen-metal angle is to 180°, the more effective is the hopping mechanism. However, such a picture assumes that the spin-orbit coupling is small enough to separate the single-particle wave functions into their spin and orbital sectors. In 5d oxides, such separation is no longer valid and the evolution of the hopping integrals as a function of the bond angle becomes more complicated. In this regard, recent work pointed towards a net increase in the overall hopping integral t between the $J_{eff} = 1/2$ electrons upon compressive strain (29).

Finally, it is worth noticing that in this material the low-energy electronic structure is largely temperature-dependent(36). Since our results are obtained at a fixed temperature of 20 K, exploring the effect of strain in the elementary excitations as a function of temperature would represent an interesting outlook for future experiments.

**Conclusions**

We have used Resonant Inelastic X-ray Scattering (RIXS) to resolve the effect of the epitaxial strain on the elementary excitations of $Sr_2IrO_4$. In particular, we have identified electron-hole pair excitations, mapped their momentum-dependence, and found that the epitaxial strain alters the energy scale of such excitations, indicating a substantial increase in the size of the Mott gap upon tensile strain. With the aid of state-of-the-art model calculations, we show that such an evolution of the Mott gap follows the ratio of the spin-exchange with respect to the hopping integral, which is responsible for the coupling between electrons (and holes) to the spin excitations in the antiferromagnetic background. In particular, under tensile strain, the collective magnetic excitations undergo a large softening at the magnetic zone boundary **Q** = (1/2,0), accompanied by hardening at **Q** = (1/4,1/4). This behavior entails an evolution from long-range anisotropic towards nearest-neighbor Heisenberg-like pseudospin interactions upon tensile strain. We suggest that the change in the Ir-O-Ir bond angle, induced by the strain, is responsible for the change in the degree of anisotropy in the magnetic interactions. However, the observed crossover from short-range to long-range interactions may also originate in a transition from Mott-like to Slater-like physics. Altogether, this suggests that the localized limit might be more appropriate to describe the system in the tensile-strain case, where the effective magnetic interactions are more akin to the 'high-$T_c$' copper oxides. Conversely, the importance of non-local interactions seems to increase upon compressive strain. Indeed, such a scenario is also corroborated by the changes in the size of the charge gap, being larger upon tensile strain. In summary, our study provides a microscopic view of the electron- and pseudospin-lattice interaction in $Sr_2IrO_4$ and demonstrates that epitaxial strain, combined with other tuning knobs such as doping or interface engineering, is a promising route towards inducing new functional properties in high-spin orbit coupled oxides.

**Materials and Methods**

Thin films of $Sr_2IrO_4$ were grown on $GdScO_3$(110) (GSO), $SrTiO_3$(100) (STO), $(LaAlO_3)_{0.3}(Sr_2AlTaO_6)_{0.7}$(100) (LSAT) and $NdGaO_3$(110) (NGO) by pulsed laser deposition to achieve an in-plane strain ε = +1.5%, +0.5% and -0.5% and -0.7%, respectively. In our convention, ε = 100*(a-$a_0$)/$a_0$ where a and $a_0$ are the strained and bulk pseudocubic in-plane lattice parameters. The epitaxial strain levels are experimentally determined using x-ray diffraction. The x-ray absorption (XAS) and Resonant Inelastic X-ray Scattering (RIXS) measurements at the O K-edge (≈530 eV) were carried out at the ADRESS beamline of the Swiss Light Source, Paul Scherrer Institut (64). The Ir $L_3$-edge (≈11.214 KeV) RIXS measurements were performed at the ID20 beamline of the European Synchrotron Radiation Facility (ESRF) (65) and at the 27 ID-B beamline of the Advanced Photon Source (APS) (66). The sample temperature was kept at 20 K during all measurements. The magnetic dispersion, measured with RIXS, was fitted with an anisotropic Heisenberg model with first-, second- and third- neighbor contributions (SI Appendix, eq. S3)(38, 52), and the exchange parameters for each sample were obtained employing least-square fitting. Theoretical predictions of the intensity and dispersion of the electron-hole pair (EHP) excitations measured in O K-edge RIXS were obtained by numerically integrating the occupied and unoccupied bands over momentum space. The dispersion of valence (conduction) bands was obtained using an extended multiplet *t-J*-like model evaluated using Green functions in the framework of the self-consistent Born approximation (60, 67). Details of the sample preparation, experimental conditions, and calculation methods can be found in the SI Appendix.

**Acknowledgments**

We gratefully acknowledge C. Sahle for experimental support at the ID20 beamline of the ESRF. The soft x-ray experiments were carried out at the ADRESS beamline of the Swiss Light Source, PSI. E. Paris and T.S. would like to thank X. Lu and C. Monney for valuable discussions. The work at PSI is supported by the Swiss National Science Foundation through project no. 200021_178867, the NCCR MARVEL and the Sinergia network Mott Physics Beyond the Heisenberg Model (MPBH) (SNSF Research Grants No. CRSII2_160765/1 and No. CRSII2_141962). K. W. acknowledges support by the Narodowe Centrum Nauki (NCN) Project No. 2016/22/E/ST3/00560 and No. 2016/23/B/ST3/00839. E.M.P. and M. N. acknowledge funding from the European Union's Horizon 2020 research and innovation programme under the Marie Sklodowska-Curie grant agreement No 754411 and No 701647, respectively. M.R. was supported by the Swiss National Science Foundation (SNSF) under Project No. 200021 – 182695. This research used resources of the Advanced Photon Source, a U.S. Department of Energy (DOE) Office of Science User Facility operated for the DOE Office of Science by Argonne National Laboratory under Contract No. DE-AC02-06CH11357.

**Figures captions**

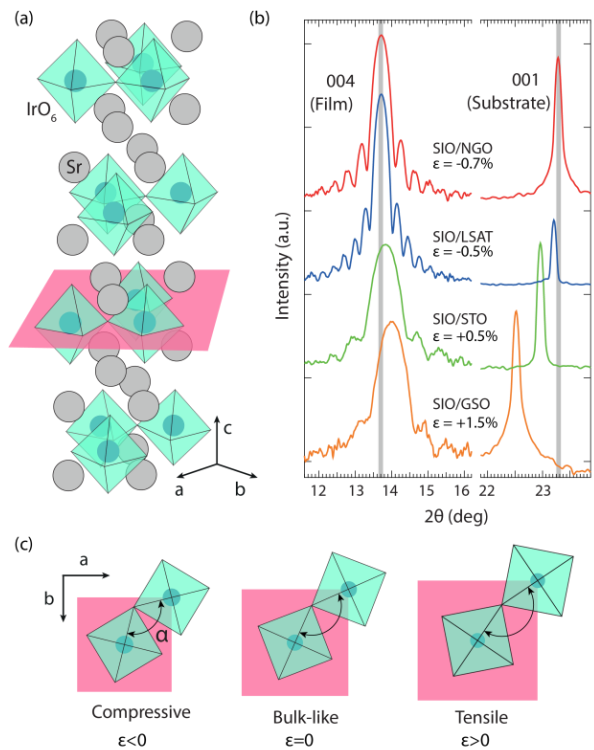

**Figure 1**. Strain engineering of Sr2IrO4. (a) Crystal structure of the single-layer perovskite Sr2IrO4. The polyhedra represent the IrO6 octahedron while the gray spheres represent the Sr ions. (b) X-ray θ-2θ diffraction scans across the (004) reflection of the Sr2IrO4 films grown on GSO (110), STO (100), LSAT (100) and NGO (110), showing the change in the unit cell induced by epitaxial strain. The pseudocubic (001) reflection of the substrate is also shown (on a different scale) for comparison. The data is presented on a logarithmic scale. (c) Pictorial view of the evolution of the local Ir-O bond in the a,b plane (cut along the plane shown in panel (a)) upon epitaxial strain, assuming that both the Ir-O bond length and Ir-O-Ir bond angle are evolving (not in scale). The box depicts the relative expansion/compression of the pseudocubic unit cell.

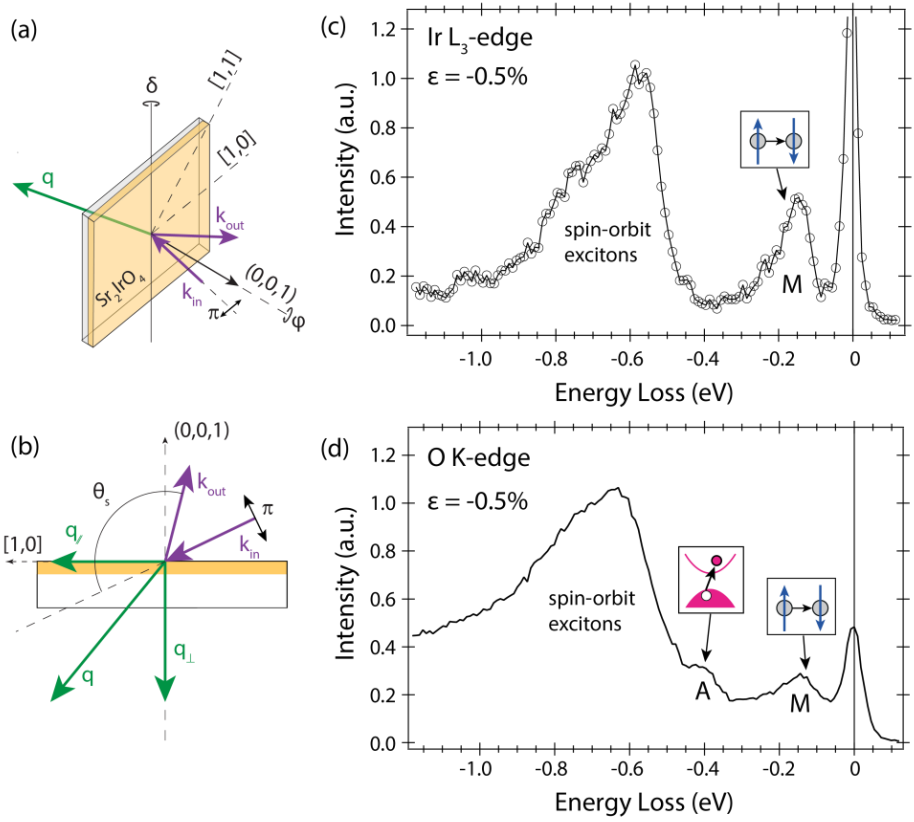

**Figure 2**. Ir L3 and O K-edge RIXS experiments. (a-b) Sketch of the experimental geometry for the RIXS experiment. The scattering angle θS is set to 130° for the O K-edge measurements and 90 ° for the Ir L3-edge measurements. (c) Ir L3-edge RIXS spectrum on a thin film of Sr2IrO4 grown on LSAT (100) with a $\varepsilon$ = -0.5% compressive strain, measured near Q = (0.25,0). (d) O K-edge RIXS spectrum on the same sample, taken at $\pi$-polarization and Q ~ (0.25,0). The pseudospin-flip mode is labeled "M" while the electron-hole pair mode is labeled "A". The intense spin-orbit exciton structure (j=3/2 to j=1/2 excitation) is also indicated. The insets show a pictorial view of the electron-hole (peak A) and pseudospin-flip (peak M) excitations.

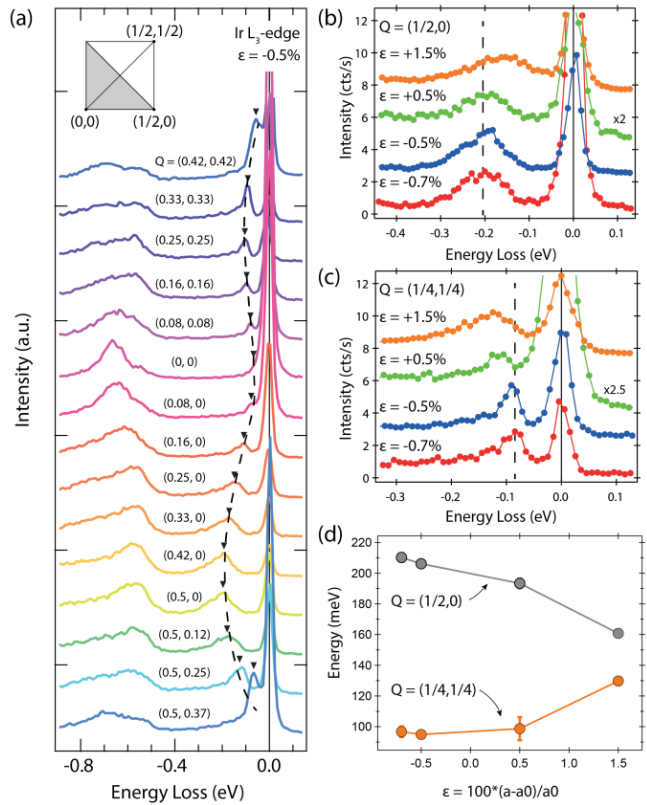

**Figure 3**. Strain and momentum-dependent Ir L3-edge RIXS. (a) Ir L3-edge RIXS spectra recorded on a thin film of Sr2IrO4 grown on LSAT(100), with $\varepsilon$ = -0.5% compressive strain for different exchanged momenta (as labeled). The inset shows the reciprocal unit cell and the AF magnetic zone (gray area). (b-c) Low energy region of the Ir L3-edge RIXS spectra at Q = (1/2,0) and Q = (1/4,1/4). Data from films grown on different substrates is shown with a vertical offset. The data for $\varepsilon$ = +0.5% is multiplied by a constant for better visualization (see label). The vertical dashed line is a guide to the eye. (d) Energy of the collective magnetic excitation at the (1/2,0) and (1/4,1/4) zone boundaries as a function of strain.

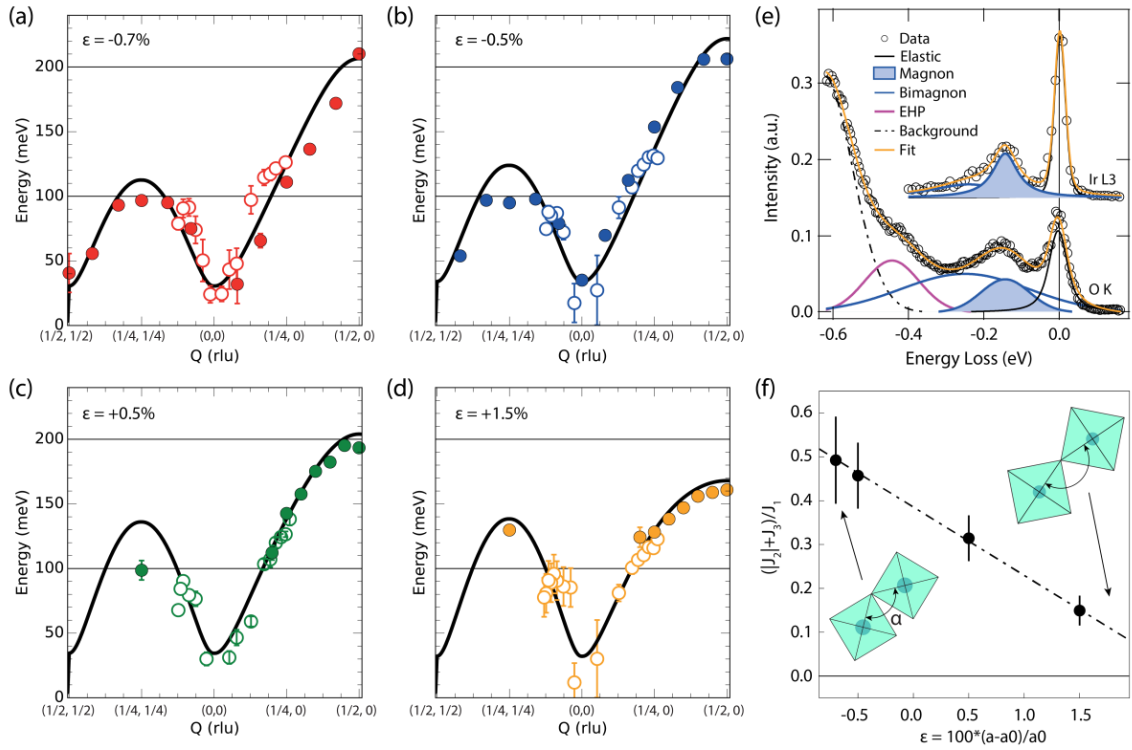

**Figure 4**. Tuning of the spin-wave mode by epitaxial strain. (a-d) Dispersion of the collective magnetic excitations in Sr2IrO4 as a function of the epitaxial strain. The data is obtained from combining the O K-edge (empty symbols) and Ir L3-edge RIXS spectra (full symbols). The black solid line is a fit to an anisotropic Heisenberg model (see text). (e) Comparison between the fitting model used for O K-edge (lower) and Ir L3-edge (upper) spectra at Q = (0.25, 0) for $\varepsilon$ = -0.5%. The data (black circles) is presented along with the best fit (orange solid line). The different fitting components are shown (see legend). (f) Strain dependence of the ratio between the magnitude of the higher-order exchange interactions (|J2|+J3) and the first nearest-neighbor exchange interaction (J1). The insets give a pictorial view of the evolution of the Ir-O-Ir bond upon strain.

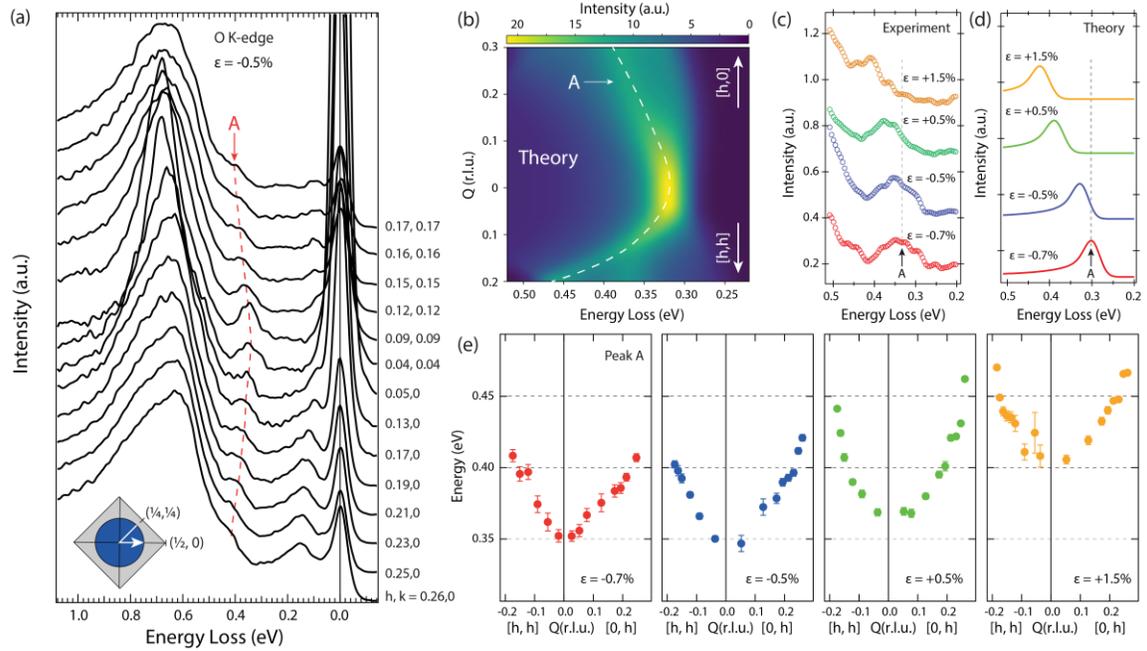

**Figure 5**. Electron-hole pair excitations from O K-edge RIXS. (a) Momentum dependence of the O K-edge RIXS spectrum of a thin film of Sr2IrO4 under $\varepsilon$ = -0.5% compressive strain, taken at 20 K with $\pi$-polarized light. The inset represents the path followed in the reciprocal space (white arrow). The gray square represents the magnetic zone of Sr2IrO4 while the blue sphere shows the maximum probing volume accessible at the O K-edge. The red dashed line marks the position of peak A. (b) Colormap representing the theoretical RIXS intensity for the EHP excitation as a function of the exchanged energy and momentum, calculated as discussed in the main text. (c) The EHP excitation from O K-edge RIXS for different strain levels, taken at Q ~ (0.04,0) r.l.u. (d) Calculated O K-edge RIXS intensity for an EHP excitation at Q = (0.04,0) as a function of strain (see main text for details). The vertical dashed line is a guide to the eye, marking the position of peak A for the highest compressive strain. (e) Energy dispersion of the EHP excitations (peak A) obtained by fitting the O K-edge RIXS spectra. Different panels represent different strain levels.

Supplementary Information for

Strain-engineering of the charge and spin-orbital interactions in $Sr_2IrO_4$

Eugenio Paris[a,1], Yi Tseng[a], Ekaterina M. Pärschke[b,c], Wenliang Zhang[a], Mary H. Upton[d], Anna Efimenko[e], Katharina Rolfs[f,g], Daniel E. McNally[a], Laura Maurel[h,i], Muntaser Naamneh[a], Marco Caputo[a], Vladimir N. Strocov[a], Zhiming Wang[j,k], Diego Casa[d], Christof W. Schneider[i], Ekaterina Pomjakushina[f], Krzysztof Wohlfeld[l], Milan Radovic[a] & Thorsten Schmitt[a,1]

[a]*Photon Science Division, Paul Scherrer Institut, 5232 Villigen PSI, Switzerland*
[b]*Department of Physics, University of Alabama at Birmingham, Birmingham, Alabama 35294, USA*
[c]*Institute of Science and Technology Austria, Am Campus 1, 3400 Klosterneuburg, Austria*
[d]*Advanced Photon Source, Argonne National Laboratory, Argonne, Illinois 60439, USA*
[e]*ESRF – The European Synchrotron, 71, avenue des Martyrs, CS 40220, 38043 Grenoble, France*
[f]*Neutrons and Muons Research Division, Paul Scherrer Institut, 5232 Villigen PSI, Switzerland*
[g]*Physikalisch-Technische Bundesanstalt (PTB) Berlin, Department 8.2 Biosignals, 10587 Berlin, Germany*
[h]*Laboratory for Mesoscopic Systems, Department of Materials, ETH Zurich, 8093 Zurich, Switzerland*
[i]*Laboratory for Multiscale Materials Experiments, Paul Scherrer Institute, 5232 Villigen PSI, Switzerland*
[j] *Key Laboratory of Magnetic Materials and Devices, Ningbo Institute of Materials Technology and Engineering, Chinese Academy of Sciences, 315201 Ningbo, China*
[k] *Zhejiang Province Key Laboratory of Magnetic Materials and Application Technology, Ningbo Institute of Materials Technology and Engineering, Chinese Academy of Sciences, 315201 Ningbo, China*
[l]*Institute of Theoretical Physics, Faculty of Physics, University of Warsaw, Pasteura 5, PL-02093 Warsaw, Poland*

[1] Corresponding authors: Eugenio Paris & Thorsten Schmitt
**E-mail**: eugenio.paris@psi.ch; thorsten.schmitt@psi.ch

**This PDF file includes:**

Supplementary text
Figures S1 to S9
Tables S1
SI References

**Supplementary Information Text**

**Sample preparation and characterization.** Thin films of $Sr_2IrO_4$ were grown on $GdScO_3$(110) (GSO), $SrTiO_3$(100) (STO), $(LaAlO_3)_{0.3}(Sr_2AlTaO_6)_{0.7}$(100) (LSAT) and $NdGaO_3$(110) (NGO) to achieve an in-plane epitaxial strain of $\varepsilon$ = +1.5%, +0.5%, -0.5% and -0.7%, respectively. In our convention, the positive (negative) sign stands for tensile (compressive) strain, calculated as $\varepsilon$ = $100*(a-a_0)/a_0$, where a and $a_0$ are the strained and bulk in-plane lattice parameters. The thin film samples were grown by Pulsed Laser Deposition (PLD), using a $SrIrO_3$ polycrystalline target ablated with a laser fluency of ~ 1 J/cm$^2$. The laser is a KrF excimer laser for the $\varepsilon$ = -0.7%, -0.5% samples while a solid-state laser was used for the $\varepsilon$ = +1.5%, +0.5% samples. The epitaxial growth conditions are achieved with an oxygen partial pressure of 10$^{-2}$ mbar while the temperature of the substrate is kept at ~950 °C. All samples show insulating behavior at the temperature of the measurement. While the sample quality is comparable in all the investigated samples, the sample thickness presents some degree of variability, ranging from 15 to 29 nm, depending on the particular substrate used.

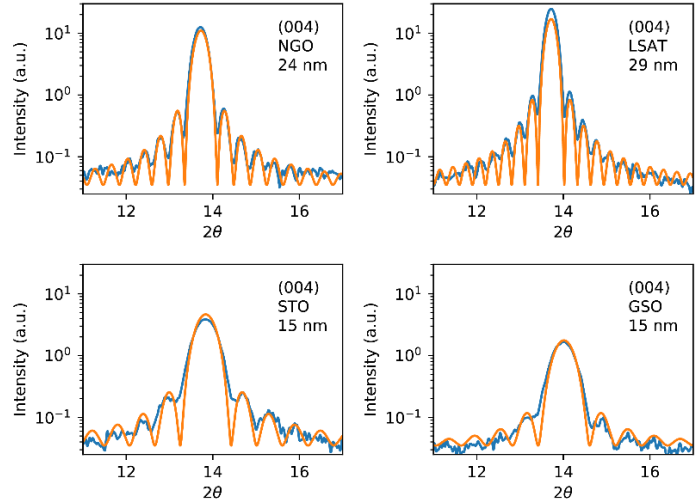

**Fig. S1**: Measured θ-2θ x-ray diffraction patterns around the 004 reflection (blue) of Sr$_2$IrO$_4$ thin films grown on different substrates, as indicated in the legend. The orange solid line represents a simulation of the finite-size fringes due to the sample thickness. The thickness of each sample, used for the simulation, is reported in the legend.

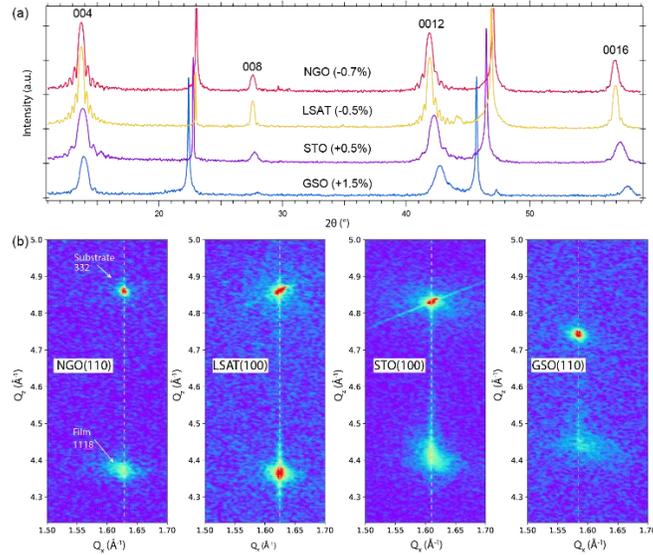

**Fig.S2**: (a) X-ray θ-2θ diffraction scans on Sr$_2$IrO$_4$ thin films grown on different substrates, as indicated. The diffraction scans show a continuous evolution of the c-axis lattice parameter as a function of strain. (b) Reciprocal space maps (RSM) around the pseudocubic 103 reflection (orthorhombic 332) of the substrates. The Q$_x$ and Q$_z$ directions indicate the in-plane and out-of-plane directions of the reciprocal vector, respectively. All data is shown in logarithmic scale, to enhance the weak diffraction peaks from the film.

Figure S1 shows the same x-ray diffraction (XRD) data as in Fig. 1 in the main text (blue line), along with the best fit result (orange line) using the equation given in Ref. (1), describing the effect of sample thickness on the XRD pattern. From the fitting, we determine the thickness of all samples: the thickness is 15 nm when the sample is grown on STO or GSO, 24 nm on NGO, and 29 nm on LSAT. The simulations explain well the difference, observed among different samples, in the visibility and spacing of the diffraction fringes around the 004 peak. Since we do not assume any kind of disorder in our simulation, the change in the peak width exhibited by our XRD data is predominantly due to finite size effects in the diffraction measurement along the out-of-plane direction.

The θ-2θ diffraction scans for the samples grown on different substrates are shown in Fig. S2(a) over a wide angle range. In order to quantify the effective strain levels achieved in our series of samples, we perform reciprocal space maps (RSM) i.e. we collect the diffraction intensity as a function of in-plane ($Q_x$) and out-of-plane ($Q_z$) reciprocal directions. As shown in Fig. S2(b), in the RSM data, the $Sr_2IrO_4$ 11$\underline{18}$ peak is visible along with the 332-reflection of the substrate (103 in pseudocubic notation). The substrate peak position shifts to lower $Q_x$ values going from compressive to tensile strain, due to the expansion of the in-plane lattice parameter. The film peaks appear at the same in-plane wave vector as the substrate, indicating a coherent epitaxial strain. The lattice parameters and misfit strain levels, evaluated by analyzing the XRD maps, are reported in Table S1.

**Table S1**. Measured lattice parameters and in-plane misfit strain for the series of $Sr_2IrO_4$ samples. The a-axis (in-plane lattice parameter) is given in the pseudocubic notation.

| Substrate  | a-axis [Å]        | c-axis [Å]         | in-plane ε [%]    |
|------------|-------------------|--------------------|-------------------|
| NGO (110)  | 3.857 (± 0.005)   | 25.812 (± 0.002)   | −0.73 (± 0.10)    |
| LSAT (100) | 3.865 (± 0.001)   | 25.805 (± 0.002)   | −0.52 (± 0.01)    |
| STO (100)  | 3.905 (± 0.002)   | 25.596 (± 0.004)   | +0.50 (± 0.06)    |
| GSO (110)  | 3.944 (± 0.010)   | 25.406 (± 0.004)   | +1.53 (± 0.25)    |

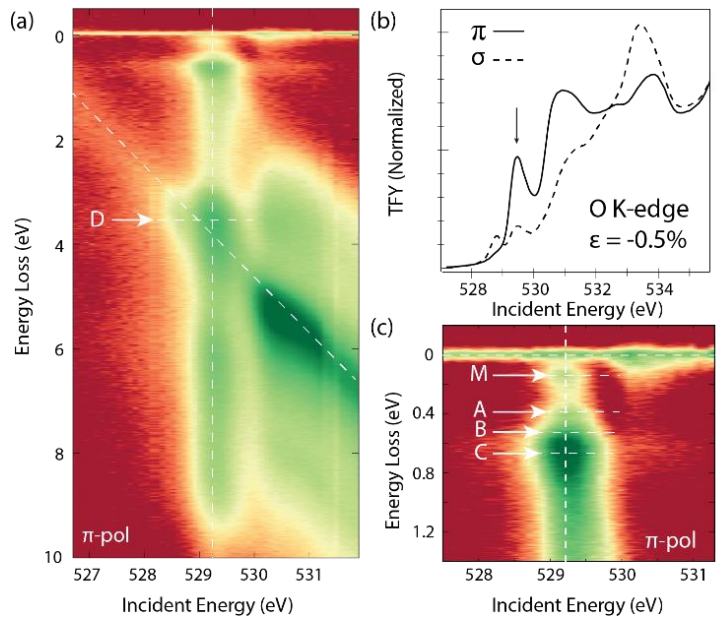

**Fig.S3**: (a) Incident energy vs energy loss RIXS map taken across the O K-edge in a $Sr_2IrO_4$/LSAT(100) thin film sample. (b) O K-edge XAS taken at grazing incidence ($\theta$ = 15°) for two orthogonal polarizations of the x-rays. The black arrow shows the incident energy used for the momentum-dependent O K-edge RIXS experiment. (c) Incident energy vs energy loss map in the low-energy region of the O K-edge RIXS spectrum. The labels in panels (a) and (c) refer to the main excitations such as magnon (M), electron-hole pair (A) and orbital excitations (B,C,D).

**XAS and RIXS measurements.** The x-ray absorption (XAS) and Resonant Inelastic X-ray Scattering (RIXS) measurements at the O K-edge (≈530 eV) were carried out at the ADRESS beamline of the Swiss Light Source, Paul Scherrer Institut (2). The high-brilliance x-ray beam (≈$10^{13}$ photons/s) was focused down to a beam spot of ≈ 4x55 µm$^2$ at the sample. In the XAS measurements, the polarized light is incident at an angle of ≈15° with respect to the film surface and the spectra are recorded by collecting the total fluorescence yield. Figure 2(a) in the main text shows a sketch of the experimental geometry for the RIXS measurements. The angle of incidence δ was varied during the experiment, allowing to span the $q_{//}$ in-plane momentum transfer across the Brillouin zone. The scattering angle was set to $\theta_S$ = 130° which allows to cover 27% and 17% of the [h,0] and [h,h] high symmetry directions in the pseudo-cubic notation with a = b ≈ 3.90 Å. The polarization chosen for this experiment was π, i.e. with the x-ray polarization vector lying in the scattering plane. The combined energy resolution of the RIXS instrument was 55 meV full width at half maximum (FWHM).

Fig. S3(b) shows the O K-edge XAS spectrum on a thin film of $Sr_2IrO_4$ grown on LSAT (100), i.e. a strain level $\varepsilon$ = -0.5%, taken with two orthogonal x-ray polarizations. The overall spectral features, as well as the large linear dichroism, are in agreement with previous experiments on single crystalline samples (3, 4), indicating the high quality of the films. XAS at the O K-edge is sensitive to final states of p-symmetry. However, the strong hybridization between O 2$p$ and Ir 5$d$ states provides sensitivity to the Ir 5$d$ unoccupied orbitals. The region between 528.5 eV and 530 eV photon energy contains transitions to the 5$d$ $t_{2g}$ whereas the region between 531 and 535 eV is sensitive to the Ir 5d $e_g$ orbitals (4). At π polarization, the peak located at 529.3 eV is sensitive to the Ir *5d* orbitals with yz+zx symmetry, hybridized via the planar oxygen $2p_x$/$2p_y$ orbitals (4).

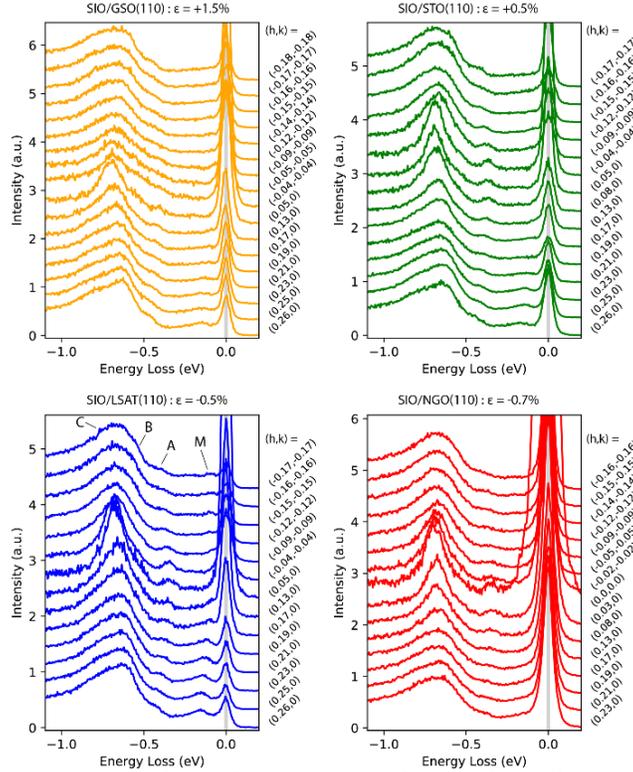

**Fig.S4**: Momentum-dependence of the O K-edge RIXS data for different strain levels, collected with 529.3 eV incoming photon energy and $\pi$-polarization at a temperature of 20 K. The RIXS spectra are normalized with respect to the high-energy fluorescence features (E> 5eV). The exchanged wave vector, expressed in reciprocal lattice units, is specified for each spectrum.

To gain insight on the resonance profile of the elementary excitations, we have recorded RIXS spectra in steps of 0.1 eV across the O K-edge. Figure S3(a) shows the characteristic energy map obtained with π-polarized light with an incident angle of ≈15° with respect to the film surface. The map shows a strong resonance at an incident energy $h\nu$ ≈ 529.3 eV, corresponding to the peak in the pre-edge region of the O K-edge XAS spectrum, representing transitions to the unoccupied upper Hubbard band. When the photon energy is tuned to this transition, a series of Raman-like excitations emerge in the RIXS spectrum. A high-energy feature (labeled D) is observed at ≈ 3.8 eV, partially overlapping in energy with a delocalized fluorescence-like feature. The latter shows a clear enhancement in the region 530 eV < $h\nu$ < 531 eV, corresponding to a broad peak in the XAS spectrum which is sensitive to Ir 5d states of $e_g$ symmetry. With a closer look to the low energy-loss region of the map, shown in panel (c) of Fig. S3, we identify a set of Raman-like excitations, labeled A, B and C and M. These modes are electron-hole pair excitations (peak A), spin-orbit excitons (B,C) and magnons (M). The Raman-like behavior of peak-A could be surprising, as one could expect a fluorescence-like behavior in the presence of an incoherent electron-hole pair excitation connected to a continuum, in apparent opposition to our assignment. However, while the electron-hole pair excitations in RIXS can carry spectroscopic information about the charge gap, their fluorescence-like versus Raman-like behavior involves hybridization between the intermediate states accessed in the RIXS process, in a rather material specific manner (5). A Raman-like behavior for an electron-hole pair mode was also observed, for instance, in the excitonic insulator $TiSe_2$ (6).

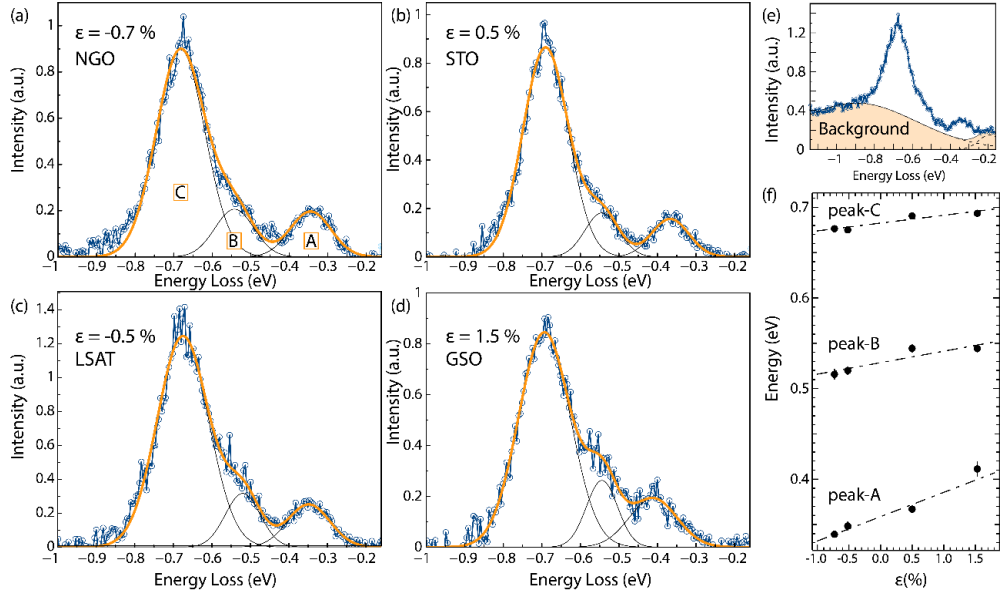

**Fig.S5**: (a-d) Background subtracted O K-edge RIXS spectra at **Q** = (0,0) for different strain levels, as specified in the legend. The solid orange line represents the best fit with three Gaussian functions, shown in black, representing the A, B and C modes. Panel (e) shows the background subtraction procedure used from the raw experimental data for $\varepsilon$ = -0.7%. (f) Energy of the A, B ad C modes as a function of the epitaxial strain.

The momentum-dependent O K-edge RIXS data, obtained with an incident energy $h\nu \approx 529.3$ eV, are shown in Fig. S4 for different strain levels. The evolution of the spin-wave and electron-hole pair excitations as a function of strain is discussed in the main text. Here, we address briefly the evolution of the B and C modes at a fixed momentum transfer. After removing the high-energy background and the magnetic spectral weight (as shown Fig. S5(e)), we have extracted the behavior of the A,B, and C modes upon strain, using Gaussian fitting. Fig. S5 (a)-(d) shows the A, B, and C excitations, measured at **Q** = (0,0), along with the best fit curves. As shown in Fig. S5(f), all three modes harden in energy upon tensile strain. In particular, peaks B and C show similar rate of change with strain, while peak-A shows a larger rate of change, compatible with a different nature of this excitation with respect to the other two. Namely, peak A is composed of electron-hole pair excitations across the Mott gap with dominant j=1/2 character (see main text), while B and C are composed by intra-$t_{2g}$ dd-excitations. However, as pointed out in previous work, all these excitations clearly disperse as a function of momentum, rendering the local picture not justified for a correct description. In fact, while their classification as dd-excitations relies on the local picture, the same set of excitations can be explained as electron-hole pair excitations between j=1/2 and j=3/2 bands across the Fermi level.

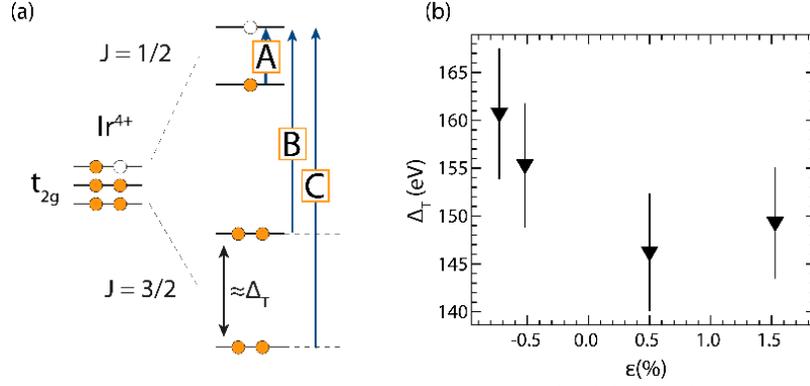

**Fig.S6**: (a) Simple schematics of the energy diagram of $Sr_2IrO_4$ and the related elementary excitations in the single-ion picture. The A transition represents the electron-hole pair excitations, while B and C are two branches of the "spin-orbit excitons", derived from intra-$t_{2g}$ dd-excitations. (b) Evolution of the $\Delta_T$ parameter, extracted from the energy of the B and C modes.

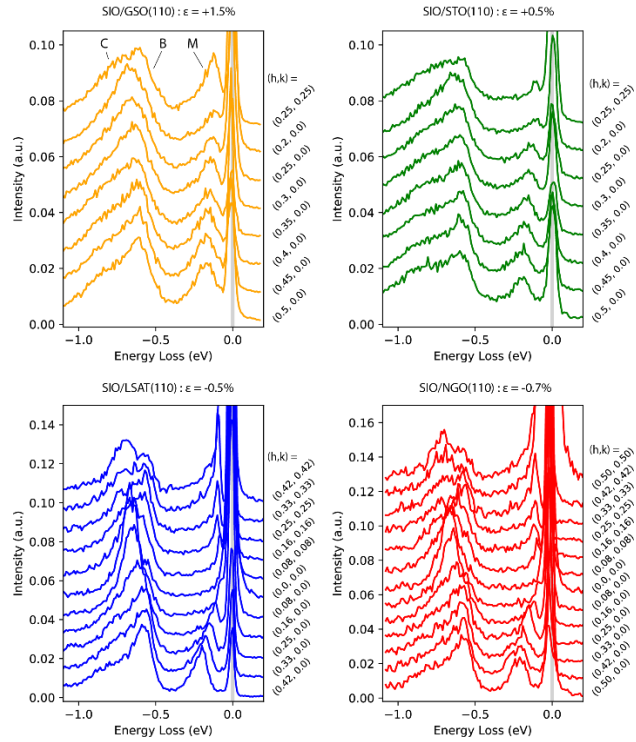

**Fig.S7**: Ir $L_3$-edge RIXS data for different strain levels, collected at a temperature of 20 K. The exchanged wave vector, expressed in reciprocal lattice units, is specified for each spectrum. The RIXS spectra are normalized by their total area.

In the following, we will focus on the effect of strain on the B and C modes at $\mathbf{Q} = (0,0)$. As shown in Fig. S6(a), within the validity of the local picture, the difference in energy between the B and C modes represents the splitting within the j=3/2 orbital manifold induced by a tetragonal distortion ($\Delta_T$). We observe that $\Delta_T$ tends to decrease by increasing the tensile strain (see Fig. S6 (b)). This behavior can be related to the evolution of the crystal structure. In the bulk $Sr_2IrO_4$ crystal, the $IrO_6$ octahedron is compressed in the in-plane direction and expanded in the out-of-plane direction. The tensile strain reduces the tetragonal distortion in the material by inducing tensile strain fields in the in-plane (⊥ c) direction and compressive strain fields in the out-of-plane (// c) direction. This result underline that a modulation of the bond lengths is likely at play upon strain.

The Ir $L_3$-edge (≈11.214 KeV) RIXS measurements were carried out at the ID20 beamline of the European Synchrotron Radiation Facility (ESRF) (7) and at the 27 ID-B beamline of the Advanced Photon Source (APS) (8). The π-polarized x-ray beams were focused on the sample using a Kirkpatrick-Baez mirror. We use a spherical diced Si (844) crystal to analyze the emitted radiation with a combined instrumental resolution of 24 meV and 35 meV FWHM at the ID20 and 27 ID-B beamlines, respectively. The data was collected with a $\theta_s$ = 90° scattering geometry to minimize the elastic scattering. Due to the large penetration depth of photons at the Ir $L_3$-edge, the x-ray beam was impinging at near grazing incidence with a glancing angle < 2° to maximize the signal from the thin films. The momentum-dependent Ir $L_3$-edge RIXS spectra are shown in Fig. S7 for different strain levels. A sharp dispersive low-energy excitation, corresponding to the spin-wave (M) is clearly resolved. In the region 500-800 meV, we observe the orbital excitations (B and C).

**Computational Methods.** We start by computing the RIXS intensity for an electron-hole interband transition, as described in the main text, Eq. (1). Thus, we first need to calculate the dispersion of the valence (occupied) and conductance (unoccupied) bands, $\epsilon_v(\mathbf{k})$ and $\epsilon_c(\mathbf{k})$ respectively. The dispersions of occupied and unoccupied states are obtained from the spectral functions as imaginary part of the single-particle Green functions of electron removal and addition states, respectively:

$$I(\mathbf{k}, \omega) = Im\,[G(\mathbf{k}, \omega)] \qquad [S1]$$

where $G(\mathbf{k},\omega)$ is the Green function describing the propagation of a single charge introduced into the ground state with commensurate electron occupancy (9, 10):

$$G = \langle AF | a_k \frac{1}{(\omega - H + i\delta)} a_k^\dagger | AF \rangle \qquad [S2]$$

Here, $a_k^\dagger (a_k)$ creates (annihilates) a charge excitation with a momentum **k** on top of the AF ground state: a hole for occupied bands, or an electron for unoccupied bands. The Hamiltonian *H* describes the propagation of the given charge excitation dressed by the low energy excitations of the ground state, i.e. magnons. The Hamiltonian *H* for either electron addition or removal states has a form of the extended *t-J*-like model, as in (9).

First, the magnetic term of the *t-J*-like model describes the dispersion of magnons created in the AF ground state. It is given by a $J_1$-$J_2$-$J_3$ Heisenberg model including an anisotropic exchange, which has recently been shown to be important for a correct description of the magnon dispersion observed in $Sr_2IrO_4$ (11):

$$H_J = \sum_{<i,j>} J_1 \left[ S_i^x S_j^x + S_i^y S_j^y + (1 - \Delta) S_i^z S_j^z \right] + \sum_{<<i,j>>} J_2 \vec{S_i} \cdot \vec{S_j} + \sum_{<<<i,j>>>} J_3 \vec{S_i} \cdot \vec{S_j} \qquad [S3]$$

The values of the exchange parameters are obtained using least-square-fitting of the experimental pseudospin-wave dispersion in strained $Sr_2IrO_4$ (see Fig. 4(a-d) in the main text) and are given in Table 1 (main text). As changing Δ in a reasonable parameter range did not show improvement of the fit, we have fixed it to the zero-strain value 0.05 for simplicity (12).

Second, the hopping term of the *t-J*-like model is derived by projecting the multi-orbital Hubbard model and expressing the orbital-dependent hopping parameters in the spin-orbit coupled basis, as described in detail elsewhere (9, 10). For bulk $Sr_2IrO_4$, these orbital-dependent hopping parameters are obtained from the density functional theory (9, 10). For the strained $Sr_2IrO_4$, the hopping parameters *t* are obtained using the following phenomenological scaling, as in (13):

$$t = t_0 e^{-\beta(\frac{a}{a_0}-1)} \quad [S4]$$

where index $0$ refers to a zero strain, $a$ stands for the in-plane lattice parameter, and $\beta$=3 (13).

At last, we calculate the Green functions in eq. S2 using the self-consistent Born approximation (9, 10). Then, once the band dispersions and their spectral weight [eq. S1] are calculated for occupied and unoccupied states, we fit the low-lying energy bands for each case with a general tight-binding model to obtain an analytical band dispersion:

$$\epsilon(q_x, q_y) = t_1 \cos(q_x)\cos(q_y) + t_2 \cos(2q_x)\cos(2q_y) \\ + t_4 \cos(4q_x)\cos(4q_y) + t_6 \cos(6q_x)\cos(6q_y) + \mu \quad [S5]$$

where $\mu$ here stands for the chemical potential. For the unoccupied states, we fit the dispersing lowest energy states with a single band $\epsilon_C$. For occupied states we consider the lowest-lying band, characterized by a singlet flavor of the electron removal state.

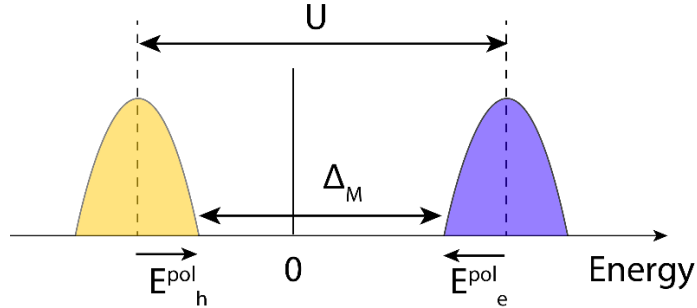

**Fig. S8**: Schematic representation of the connection between the Mott gap and the polaron binding energies. Yellow (blue) areas show the bandwidth of magnetic polarons, formed by holes (electrons) for the lower (upper) Hubbard bands. The bandwidth of polarons is defined by the strength of their coupling to the magnetic background.

Noticeably, all the parameters used in the calculation come from first principles or direct comparison to experiments, apart from the Fermi level position in the electron addition spectral function. As the chemical potential is not included in a $t$-$J$-like model, it is estimated by comparing the electron removal spectral function to a ARPES experiment (14). For the unoccupied bands, we also fix the position of the chemical potential throughout all calculations, however, we consider the zero-strain value an adjustable parameter as there is no available experimental inverse photoemission data to compare it to. Figure S9 (e-h) shows the results of the calculation of N($\omega$,**Q**), representing the electron-hole excitations across the energy gap in $Sr_2IrO_4$ as a function of strain. In particular, the conduction band $\epsilon_c(\mathbf{k})$ and the valence band $\epsilon_v(\mathbf{k})$ dispersions for different strain levels are represented in panels (a)-(d) as solid and dashed lines, respectively. By scaling the hopping parameters using eq. S4, and adapting the exchange coupling parameters from fitting the experimental magnetic dispersion for various strain levels [Fig. 4 (a)-(d) and Table 1 in the main text], we observe an increase in the separation between valence and conduction bands, hence an increase in the size of the band gap. The increase in the size of the Mott gap ($\Delta_M$) upon tensile strain can be understood by noting that

$$\Delta_M = U - E_{pol}^h - E_{pol}^e \quad [S6]$$

where $E_{pol}^h$ ($E_{pol}^e$) are binding energies for a "magnetic polaron", composed by a single hole (a single electron) added to the Mott insulating ground state with commensurate filling.

This means that the increase in $\Delta_M$ upon tensile strain can be caused either by an increase in the Hubbard U or by a decrease of the polaron binding energies, as shown schematically in Fig. S8. In our calculations, we assumed that the Hubbard U should not change with the strain, i.e. the zero-energy reference is the same for all strain levels. Such an assumption is justified, as strain affects mostly the electronic hopping and not the screening of the local Coulomb repulsion.

The change in the Mott gap is due to the modification of the polaron binding energies, which is induced by the renormalization of the hopping parameters as well as the exchange couplings. The binding energy of magnetic polarons in both valence and conductance bands depends on the ratios of the spin-exchange parameters J with respect to the hopping parameters t (15). The latter effect is well-observed also in the standard t-J model [see, e.g. Table II of Ref. (15)] and can be understood by realizing that a decrease in J/t decreases the coupling between the hole and the magnon and thus it makes the hole less bound to the magnons, lowering the polaron binding energy.

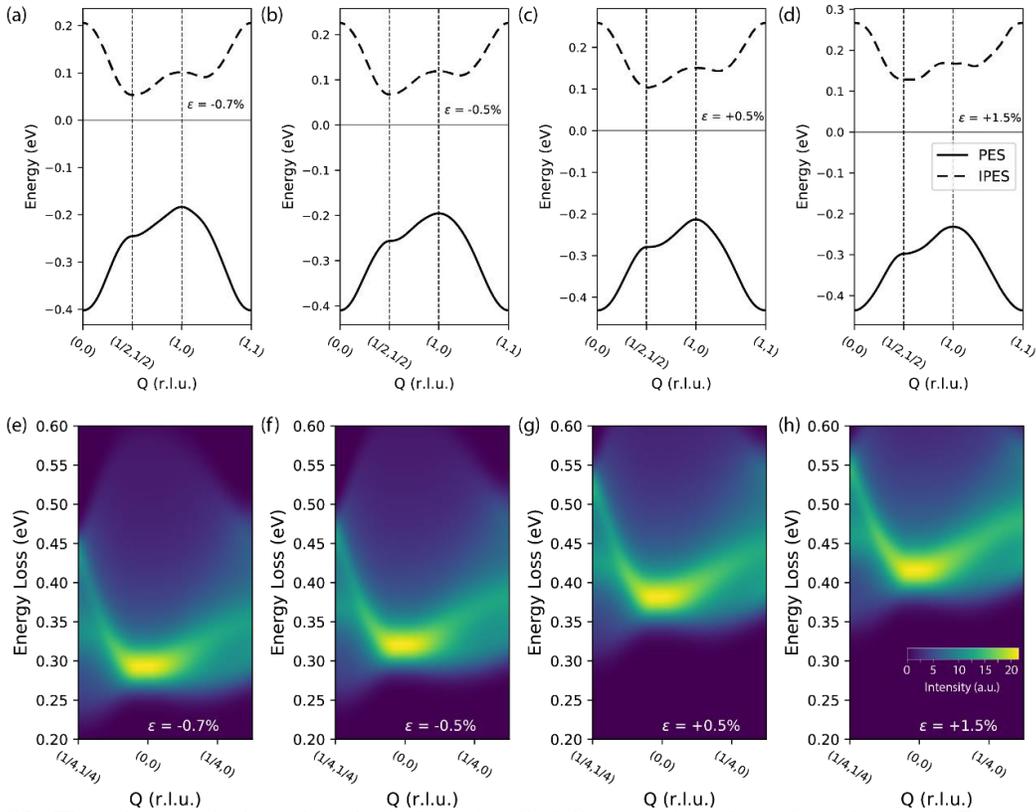

**Fig.S9**: The electron-hole pair calculations for $Sr_2IrO_4$ as a function of strain. (a-d) The valence band $\epsilon_v$ and the conduction band $\epsilon_c$ along high symmetry lines of the reciprocal space are plotted as solid and dashed curves, respectively. (e-h) The calculated N ($\omega$,Q), representing the EHP excitations, obtained for the different strain levels along the high symmetry lines, in a region of momenta accessible by O K-edge RIXS. The strain values considered are reported in the legend.

In $Sr_2IrO_4$, the hopping parameters t decrease monotonically with the tensile strain according to eq. [S4]. Experimentally obtained second- and third- neighbor spin-exchange parameters are, however, decreasing significantly quicker with tensile strain. The second- and third- neighbor spin-exchange are contributing most to the width of the polaron dispersion, since the motion of the polaron via first neighbor spin-exchange is quite confined due to stronger coupling to the antiferromagnetic background(9). The width of the polaron dispersion, in turn, defines the binding energy of the polaron, as shown in Fig. S8. Thus, the decrease in the ratio of J/t's upon tensile strain leads to lower binding energies for the magnetic polarons, which explains the observed increase of the charge gap upon the tensile strain.

Panels (e-h) in Fig. S9 represent the calculated N($\omega$,**Q**) for the corresponding strain levels, to be compared with the EHP excitations observed experimentally. In good accordance with the O K-edge RIXS experiment, our calculation reveals a hardening of the EHP excitations as a function of tensile strain.